\documentclass[%
	submission,
	final,
	notitlepage,
	narroweqnarray,
	inline,
	twoside,
	]{ieee}

\newcommand {\Xmax}{$X_{max}\,$}
\newcommand {\m}{$m$}
\newcommand {\soft}{\textquotedblleft soft\textquotedblright \,}
\newcommand {\hard}{\textquotedblleft hard\textquotedblright \,}
\newcommand {\semi}{\textquotedblleft semi-hard\textquotedblright \,}

\usepackage{verbatim}

\begin{document}

\title[Comparison of EPOS and QGSJET-II in EAS Simulation using CORSIKA ]{%
       Comparison of EPOS and QGSJET-II in EAS Simulation using CORSIKA }

\author[Chabin Ch. Thakuria]{Chabin Ch. Thakuria,
      \authorinfo{C.\,C.\,Thakuria (Corresponding author) is with the Department of Physics, Tihu College, Tihu, Nalbari, Assam, INDIA.
       Phone: $+$91\,94350\,15996, e-mail: chabin27@yahoo.com}%
\and{}and K.\ Boruah\ 
      \authorinfo{K.\ Boruah is Professor in the Department of Physics, Gauhati University, Guwahati, INDIA.
       Phone: $+$\,91\,94355\,43920,  
       e-mail: kalyaneeboruah@gmail.com}
}


\titletext{ }
\ieeecopyright{/\$0.00 \copyright\ }
\lognumber{xxxxxxx}
\pubitemident{S}
\loginfo{Manuscript received.}
\firstpage{1}

\confplacedate{ }

\maketitle

\begin{abstract} 
In this work we compare the predictions of two representative hadronic interaction models,
EPOS 1.99, and QGSJET II-03 with several extensive air showers (EAS) parameters for 
proton and iron primaries in the energy range $10^{17}$ - $10^{19} eV$ using CORSIKA-6990. 
The EAS parameters viz. depth of shower maximum, shower size, size of muon shower, 
muon number distribution, electron number distribution, 
size of hadron shower, hadron energy sum, electron muon correlations,
and, hadron energy spectra are studied in this paper.     
\end{abstract}


\section{Introduction}

         High-energy cosmic rays enter into the Earth's atmosphere resulting cascades of secondary particles known 
as extensive air showers(EAS). Information regarding the shower generating primary   particle have to be derived 
from the registered information of secondary particles at observation level. The interpretation of properties of primary 
radiation derived from air shower measurements depends on the understanding of the complex processes of high-energy 
interactions during the development of air showers~\cite{Apel}. From the number and distribution of various ground 
particles of the EAS, the reconstruction of energy and the mass of the primary particle can be done. But to relate the 
observables to primary energy and mass, more reliable algorithms and detailed air shower simulations are needed. By 
comparing the predictions from simulation with measurements one can draw conclusions on the primary mass composition of 
the arriving particles.

Again predictions from simulations suffer from systematic uncertainties mainly due to statistical fluctuations involved 
in large-scale experiments and due to modeling of HE interactions. While the electro-weak interaction processes are 
reasonably well understood; above the attained energy by the man-made accelerator, modeling of hadronic multi-particle 
production is subject to large theoretical uncertainties~\cite{D d'Enterria}. Estimation of these uncertainties are 
further difficult task.

Different hadronic interaction models predict different lateral shapes and different number of particles at observation 
level. Hence, it is possible to test and compare the available interaction models by studying these EAS parameters.
Moreover, some of these parameters like the muon content, depth of shower maximum \Xmax, the expected lateral shape etc. 
also depend on the mass of primary cosmic rays. Heavier primaries lead, on average, to a flatter distribution and lower 
value of \Xmax  and more muons.

\section{EPOS 1.99 and QGSJET II-03}

 CORSIKA ~\cite{corsika-old,corsika_upgrade} is a detailed Monte Carlo program to simulate the 4-Dimensional evolution 
of EAS in the atmosphere initiated by hadron, photon or any other particle. In CORSIKA-6990~\cite{corsika}, there are 
seven HE models, namely  DPMJET 2.55  ~\cite{dpmjet}, EPOS 1.99 ~\cite{epos,eposa}, {\sc{ne}}X{\sc{us}} 3.97 ~\cite{nexus},
QGSJET-01C ~\cite{qgsjet}, QGSJETII.3 ~\cite{qgsII,qgsII3}, SIBYLL 2.1 ~\cite{sibyll,sibyll_2.1}, and, VENUS 4.12 ~\cite{venus},
from which we can choose one model at a time for EAS simulation.\\

 To explain hadron-hadron/nucleus-nucleus collisions above the energy attainable by the man made accelerators, reliable 
and consistent hadronic interaction models are to be adopted. DPMJET, EPOS, QGSJET, and, SIBYLL are based on the Gribov's 
Reggeon approach ~\cite{gribov1,gribov2}, of Pomeron exchange in multiple scattering. The Pomeron corresponds to microscopic
partons (quarks \& gluons) cascades, can be classified into \soft, \semi, and  \hard Pomerons. Soft non-perturbative interactions 
involving large impact parameters and slow energy rise are described as \soft Pomeron exchanges, and are dominant at 
relatively low energy domain giving important contributions to total, inelastic, and diffractive cross sections.
However, in the high energy regime \semi Pomeron exchange, and, \hard Pomeron exchange describing hard interactions (with 
high energy rise) at large impact parameter, and at small impact parameters respectively are dominant.

 \subsection{QGSJET II-03}
QGSJET~\cite{qgsjet} is an addition of Gribov’s Reggeon approach~\cite{gribov1,gribov2} of hadronic and nuclear collisions
to the Quark-Gluon String model of high energy interactions. It has been generalized to treat nucleus-nucleus interactions 
and \semi processes using the so-called \semi Pomeron approach.~\cite{qgsII,qgsII3,ostapchenko}. The \semi processes can 
be described by enhanced Pomeron diagrams and proved to be of extreme importance for a correct treatment of very high energy
hadronic interactions ~\cite{ostapchenko1}. QGSJET scheme is based on the assumption that individual Pomeron exchanges occur
independent of each other, which is not true at high energy regime. At high energies parton cascades strongly overlap and 
interact with each other. These effects can be described as Pomeron-Pomeron interactions. This non-linear interaction effects
are incorporated into the  QGSJET-II-3 model, which is based on the assumption that corresponding effects are dominated by 
\soft partonic processes.
 Re summation of essential enhanced contributions corresponding to particular final states of the interaction from uncut 
diagrams (representing the elastic scattering), and from various unitarity cuts of enhanced Pomeron diagrams for all orders,
yields the final state. QGSJET II-03 is based on the obtained solutions, explicitly treating the corresponding effects in 
individual hadronic (nuclear) collisions \cite{qgsII3,ostapchenko1}.

\subsection{EPOS 1.99}
EPOS~\cite{epos} is a newly developed model emerging from VENUS~\cite{venus} and {\sc{ne}}X{\sc{us}} models~\cite{nexus}.
EPOS is a parton model, with many binary parton-parton interactions, each one creating a parton ladder. EPOS is a quantum 
mechanical model of multiple scattering approach based on partons and strings. Both the process of the particle production,
and the process of cross section calculations are consistent with the conservation of energy in EPOS. However
for other models energy conservation is not considered for cross section calculations ~\cite{hladik}.
In EPOS 1.99 reduction of the proton-nucleus cross section is done for better correlation between the number of muons
and the number of electrons at ground based air shower measurement ~\cite{eposb,eposc}.

\subsection{FLUKA 2011} 
In CORSIKA apart from seven high energy interaction models there are three low energy models, namely GHEISHA~\cite{gheisha}, 
UrQMD~\cite{urqmd}, and FLUKA~\cite{fluka}. The Fluka hadron-nucleon interaction models are based on resonance production and
decay below a few $GeV$, and on the Dual Parton model above. Two models are used also in hadron-nucleus interactions~\cite{fluka1,fluka2}.

\begin{center} 
\begin{figure}[h]
\includegraphics[width=0.5\textwidth]{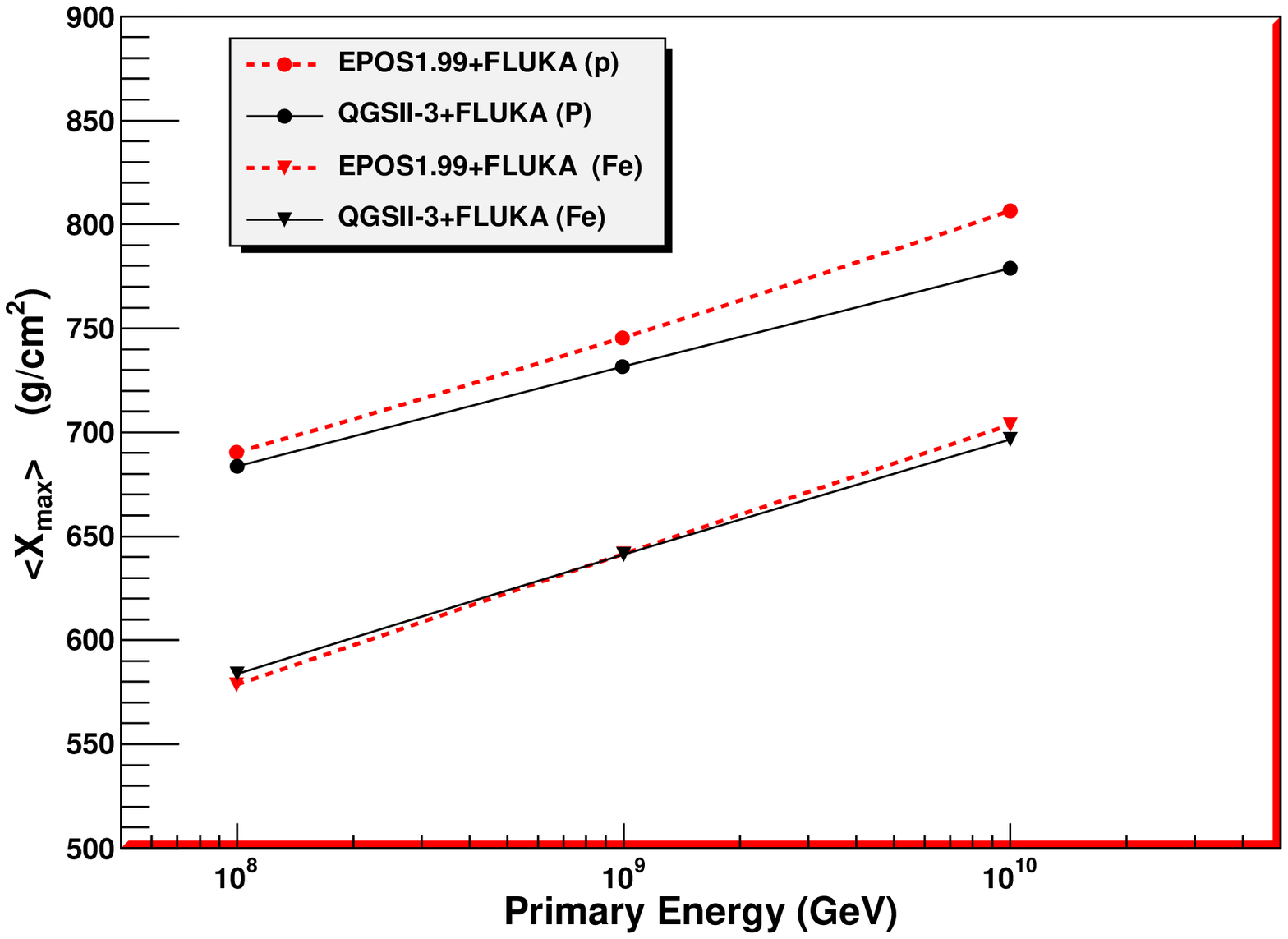}
\caption{Average $X_{max}$  Vs log of primary energy}

\label{X_max}        
\end{figure}
\end{center}

\begin{center}
\begin{figure}[h]
\includegraphics[width=0.5\textwidth]{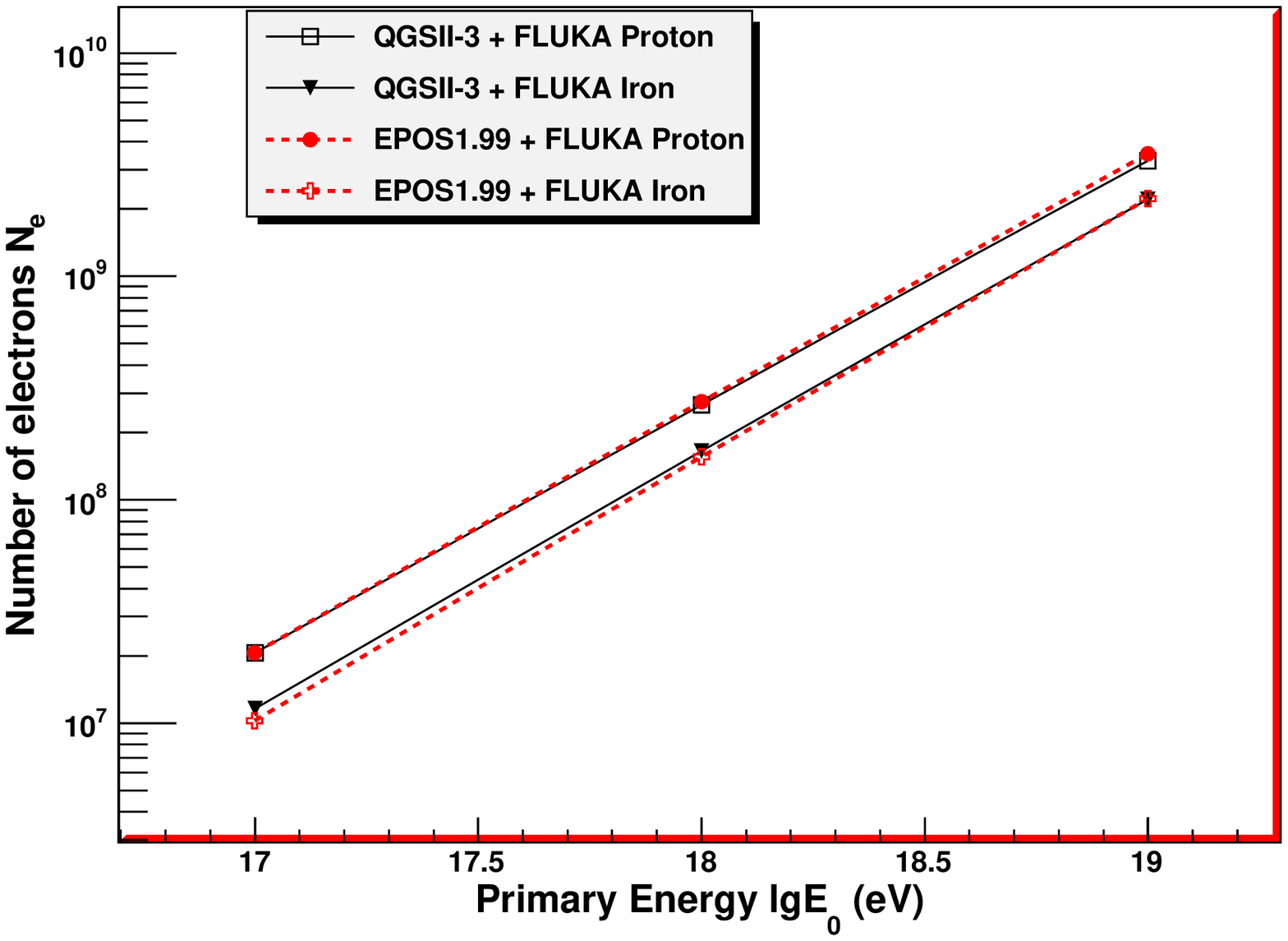}
\caption{Average electron shower size Vs log of primary energy}
 \label{Ne}        
\end{figure}
\end{center}

\subsection{Moun Number}

Muons are produced mainly by decay of charged pions and kaons in a wide energy range. Usually they are not produced directly
on the shower axis. Multiple Coulomb scattering occurred in the atmosphere and in the shielding of the detector may change
the initial direction of the muon. It is known that the reconstruction of the longitudinal development of the muon component
provides the information similar to that obtained with the fluorescence technique, but in the energy range above that accessible
by the detection of fluorescence light \cite{Doll}. Thus muon component of EAS provides a powerful key for primary mass measurement
and as well as provides information regarding hadronic interactions. In some experiment like KASCADE, truncated muon size is 
calculated by integrating muons between 40\m \, and 200\m . Instead of total number of muons, truncated muon size is considered 
for EAS study~\cite{grande}. The range of muon truncation is from 140\m-360\m \, in KASCADE Grande.

\begin{center}
\begin{figure}[h]
\includegraphics[width=0.5\textwidth]{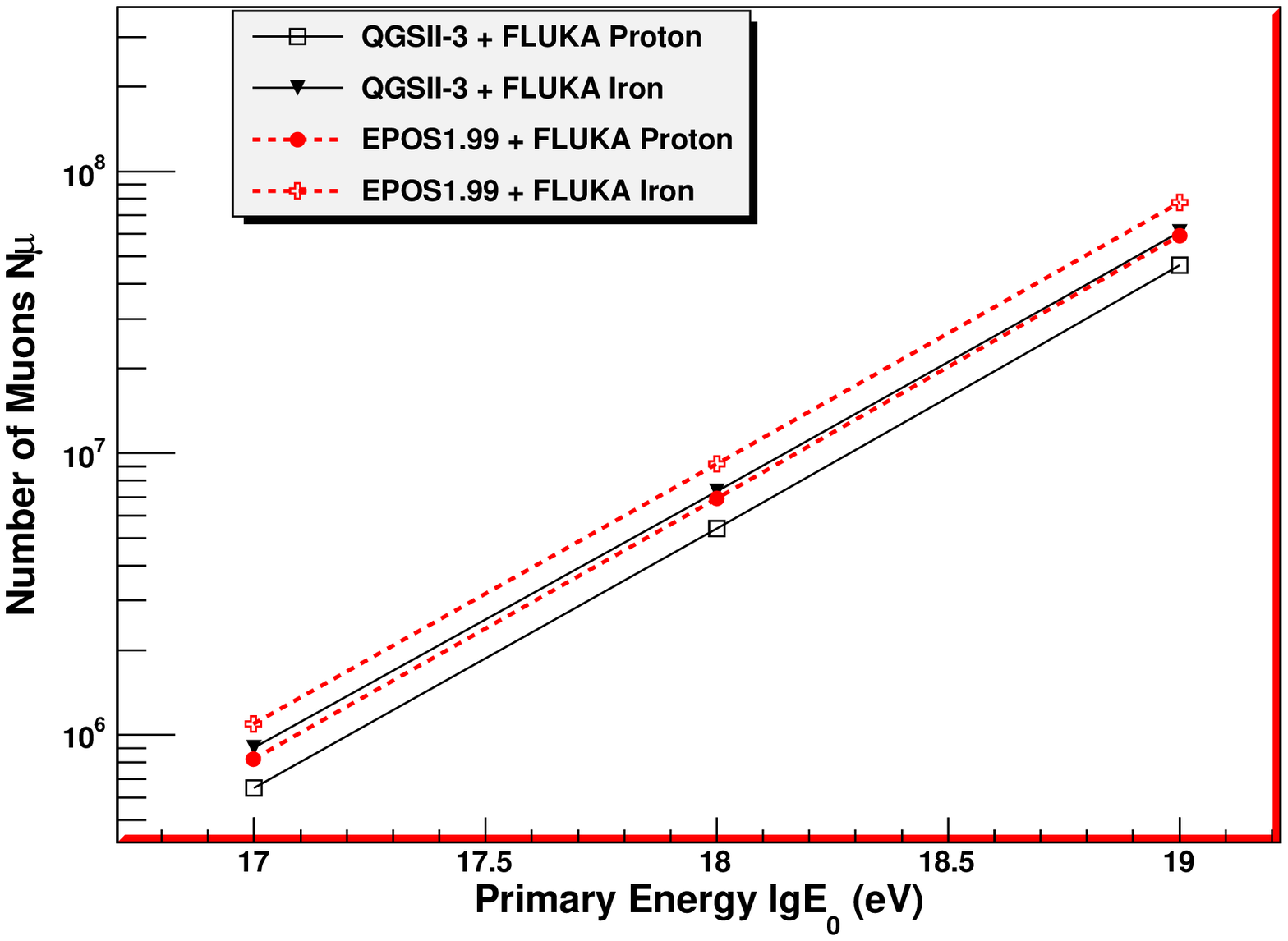}
\caption{Average muon number Vs log of primary energy}
 \label{Nmu}        
\end{figure}
\end{center}

\begin{center}
\begin{figure}[h]
\includegraphics[width=0.5\textwidth]{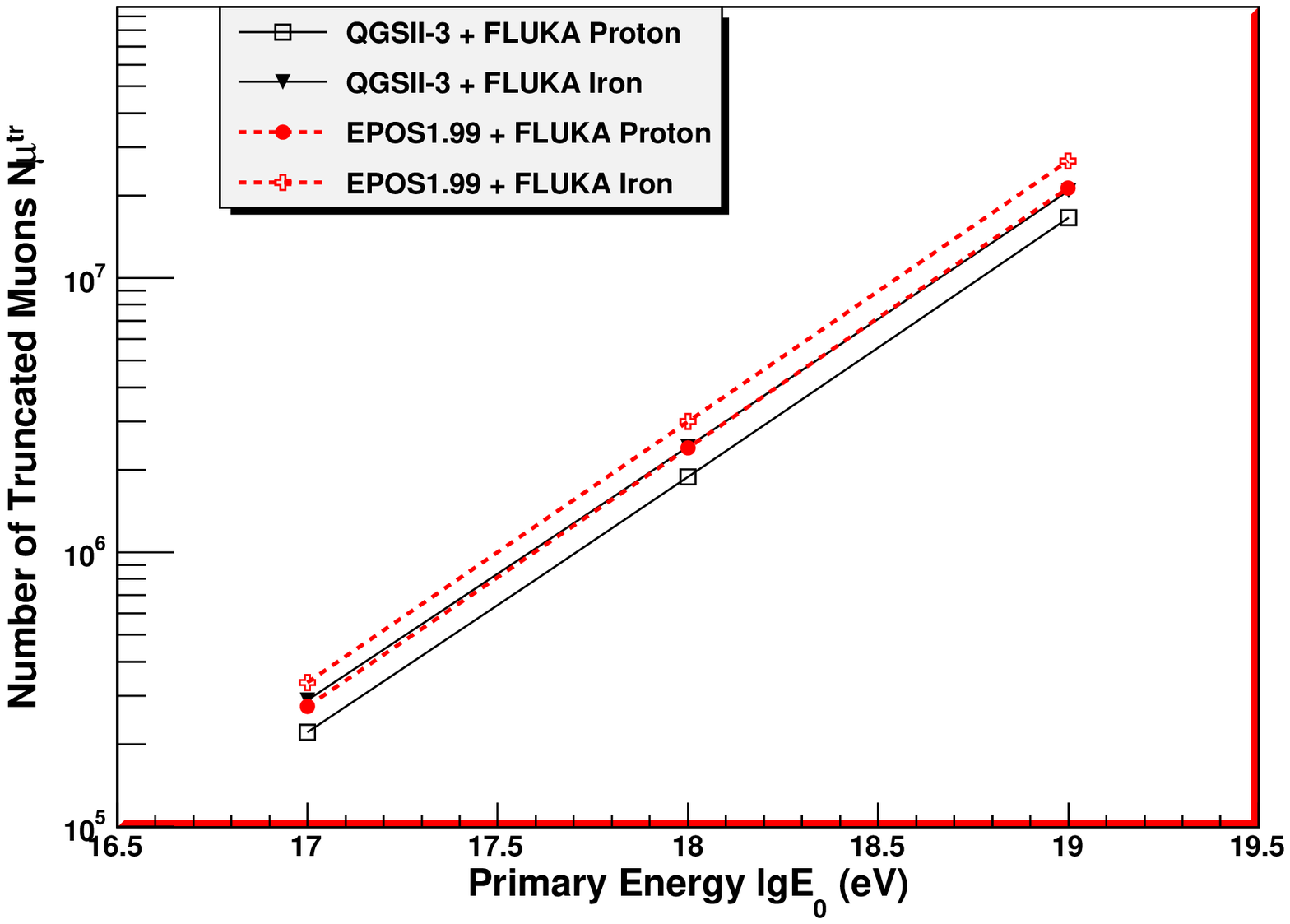}
\caption{Average truncated muon number Vs log of primary energy}
\label{Ntrmu}        
\end{figure}
\end{center}

\subsection{Depth of Shower max}

The depth of shower  maximum contains the information about the mass of the primary CR initiating the shower as well as 
about the properties of hadronic interactions involved in the process of cascade evolution. The average value \Xmax depends
on the primary energy E and on the number of nucleons A of the primary as given in the Eq.1,

  $ X_{max}$ = $\alpha$ ($ ln$ E - $ln$ A ) + $\beta$  \hfill                 (1)

  where $\alpha$ and $\beta$ depend on the the details of hadronic interactions so far as a fixed primary is considered. 
Their values are very sensitive to changes in cross-section, multiplicity and elasticity~\cite{Ulrich}.  Eq.1 can be derived
from  the simple generalized Heitler model of cascade formation due to hadronic primaries, but it is in good agreement with 
the description of the \Xmax  evolution predicted by hadronic models currently in use. Eq.1 can be expressed  as,

$ X_{max}$ = $ER_{10}$ ($ lg$ E - $lg$ A ) + $X_{init}$  \hfill                 (2)

  where $ER_{10}$ is known as elongation rate and $X_{init}$ is the depth of first interaction ~\cite{Grieder}.
  Another sensitive parameter is ${\sigma}_{X_{max}}$, expressing quantitatively the shower to shower fluctuations of \Xmax. 
It depends mainly on the cross section and less strongly on the elasticity. This makes fluctuations in \Xmax, a good parameter 
to study hadronic cross sections at ultra-high energies.\cite{Ulrich}

\section{Simulation}
The shower simulations are performed using CORSIKA-6990. Hadronic interactions at low energies(E $<$ 80 $GeV$) are modelled 
using the FLUKA-2011 code. High Energy interactions are treated with EPOS 1.99 and QGSJET II-3. Vertical showers initiated 
by primary protons and iron nuclei are simulated. Observation level is taken at the sea level. US standard atmosphere and
default magnetic field in CORSIKA are taken. 1000 showers are simulated each for three primary energies 
$10^{17}$ , $10^{18}$, and $10^{19} eV$, two primary masses(p \& Fe) and for two HE models EPOS 1.99 and QGSJET II-3. 
All together 12,000 showers are generated.
     
\begin{center}
\begin{figure}[h]
\includegraphics[width=0.5\textwidth]{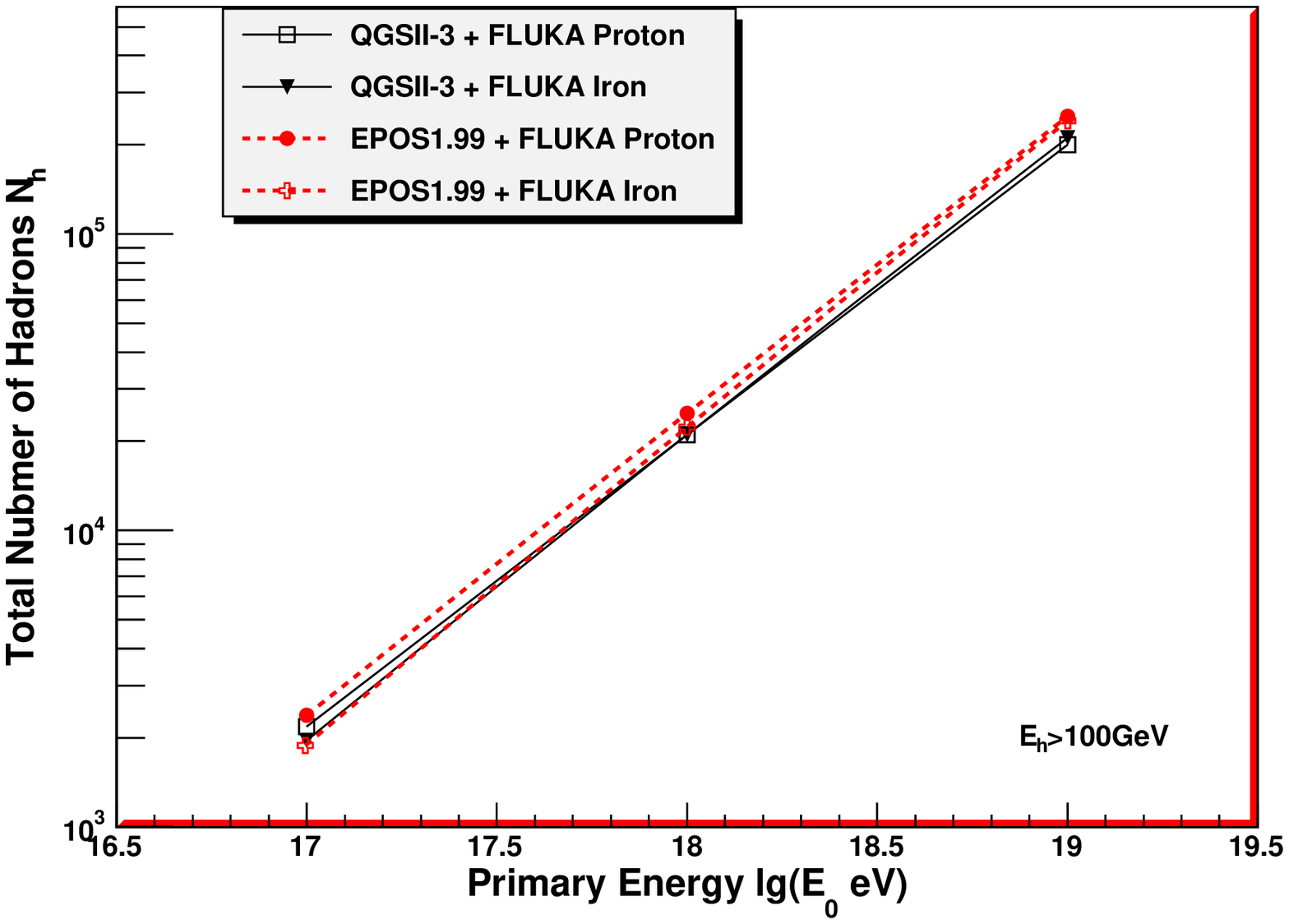}
\caption{hadron no Vs log of Primary Energy }
 \label{hadno}         
\end{figure}
\end{center}

\begin{center}
\begin{figure}[h]
\includegraphics[width=0.5\textwidth]{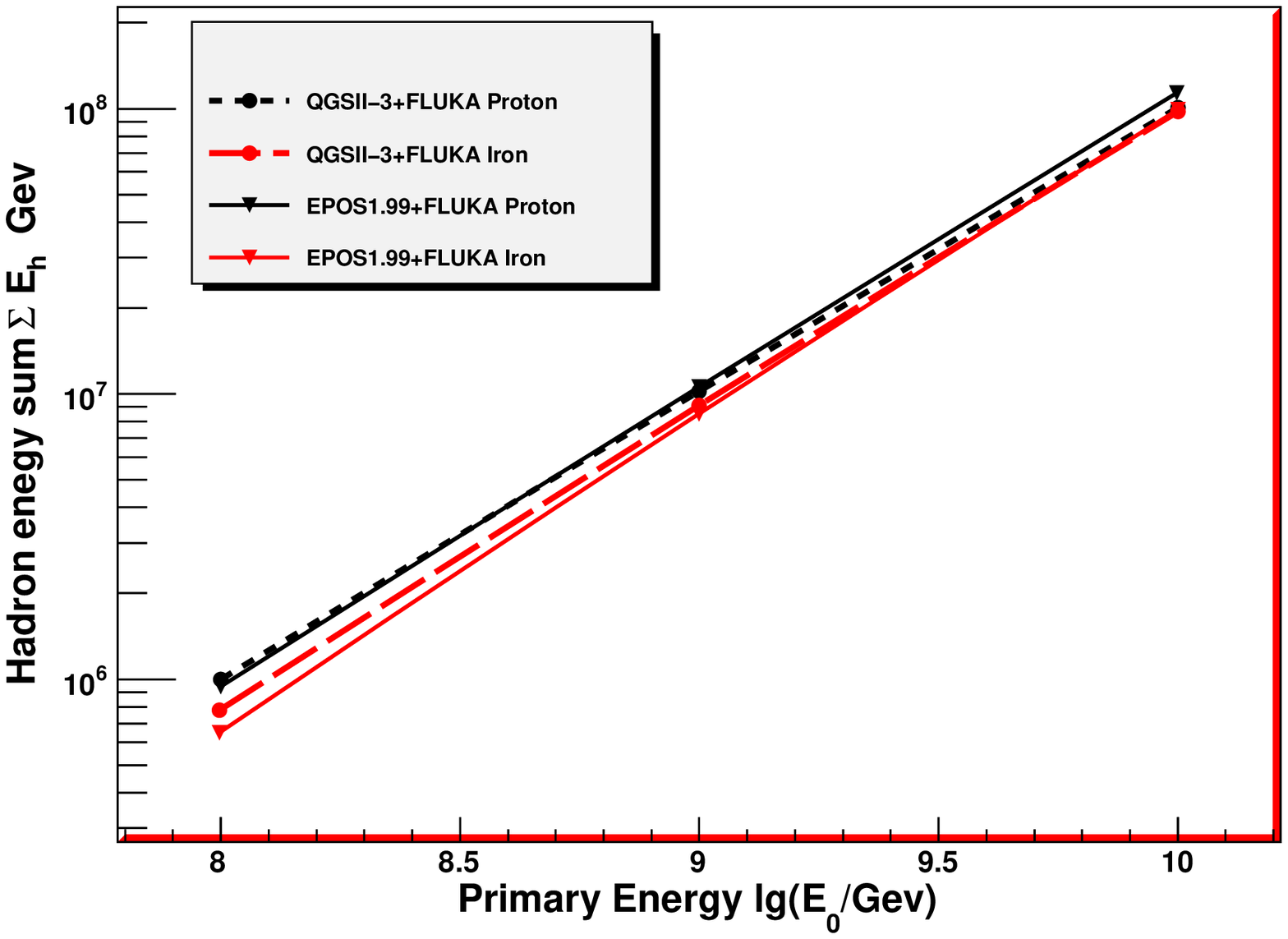}
\caption{hadron energy sum Vs log of Primary Energy }
\label{sumhad_en}         
\end{figure}
\end{center}

\begin{center}
\begin{figure}[h]
\includegraphics[width=0.5\textwidth]{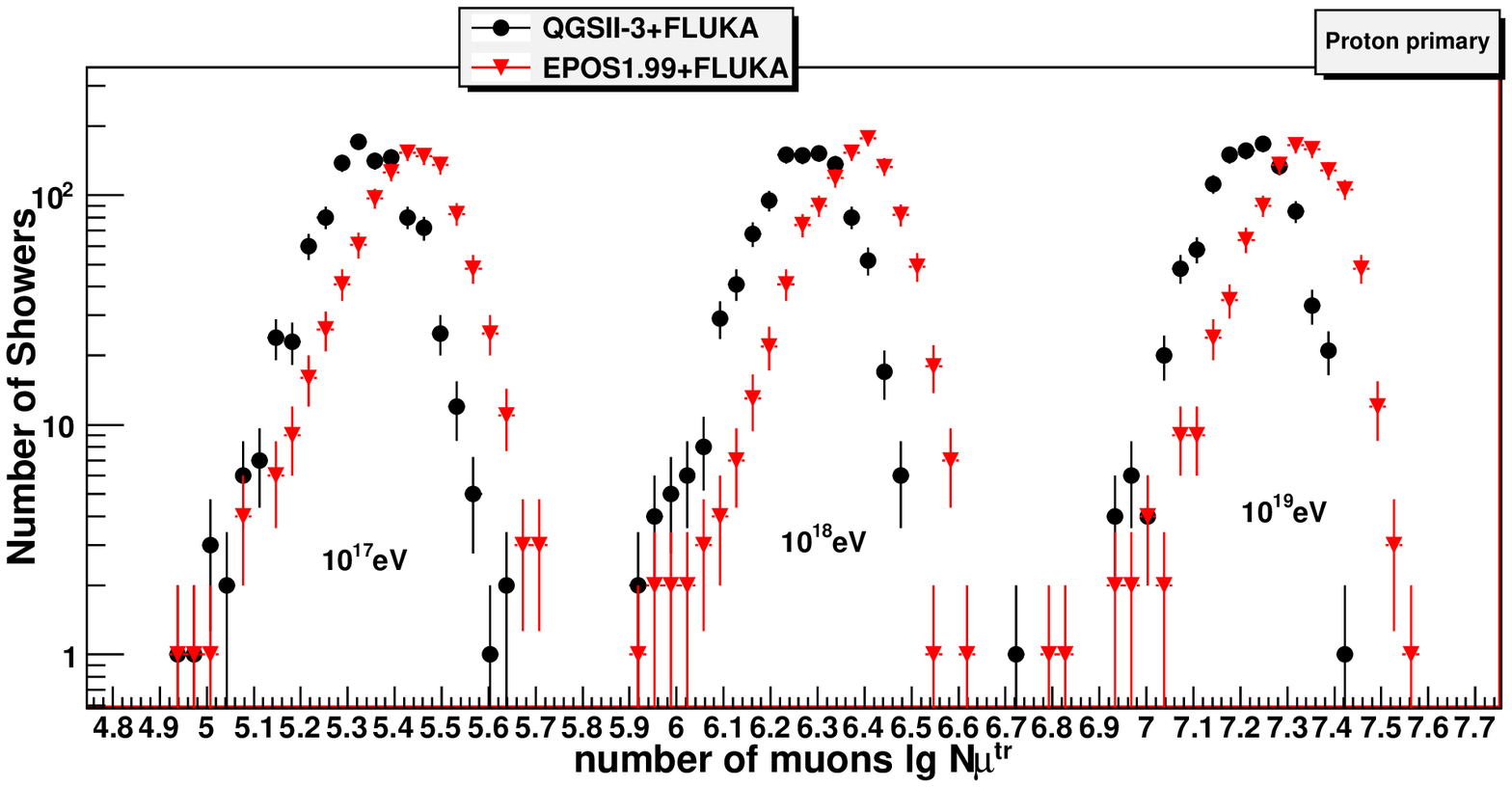}
\caption{Distribution of the number of truncated muons for proton primary }
 \label{Mu_Pr}         
\end{figure} 
\end{center}

\begin{center}
\begin{table}
\caption{Slope of \Xmax   and Intercept }
{\small
\resizebox {.5\textwidth }{!}{
\begin{tabular}{l|rl|rl}\hline
Model&\multicolumn{2}{c}{Slope of \Xmax}&\multicolumn{2}{c}{Intercept}\\\cline{2-5}
 \& Primary&\multicolumn{1}{c}{$ ER_{10}$}&$error$&\multicolumn{1}{c}{$X_{init}$}&$error$\\\hline
$EPOS-p$&   58.3690&   $\pm$1.64891&    -303.488&    $\pm$29.7108\\
$QGSII-p$&  47.5875&   $\pm$0.24876&    -125.219&    $\pm$4.48227\\
$EPOS-Fe$&     62.9539&   $\pm$1.55497&    -491.491&    $\pm$28.0169\\
$QGSII-Fe$&    55.2977&   $\pm$0.23593&    -354.504&    $\pm$4.25079 \\\hline
\end{tabular}
}}
\label{xmaxtable}
\end{table}
\end{center}

\begin{center}
\begin{figure}[h]
\includegraphics[width=0.5\textwidth]{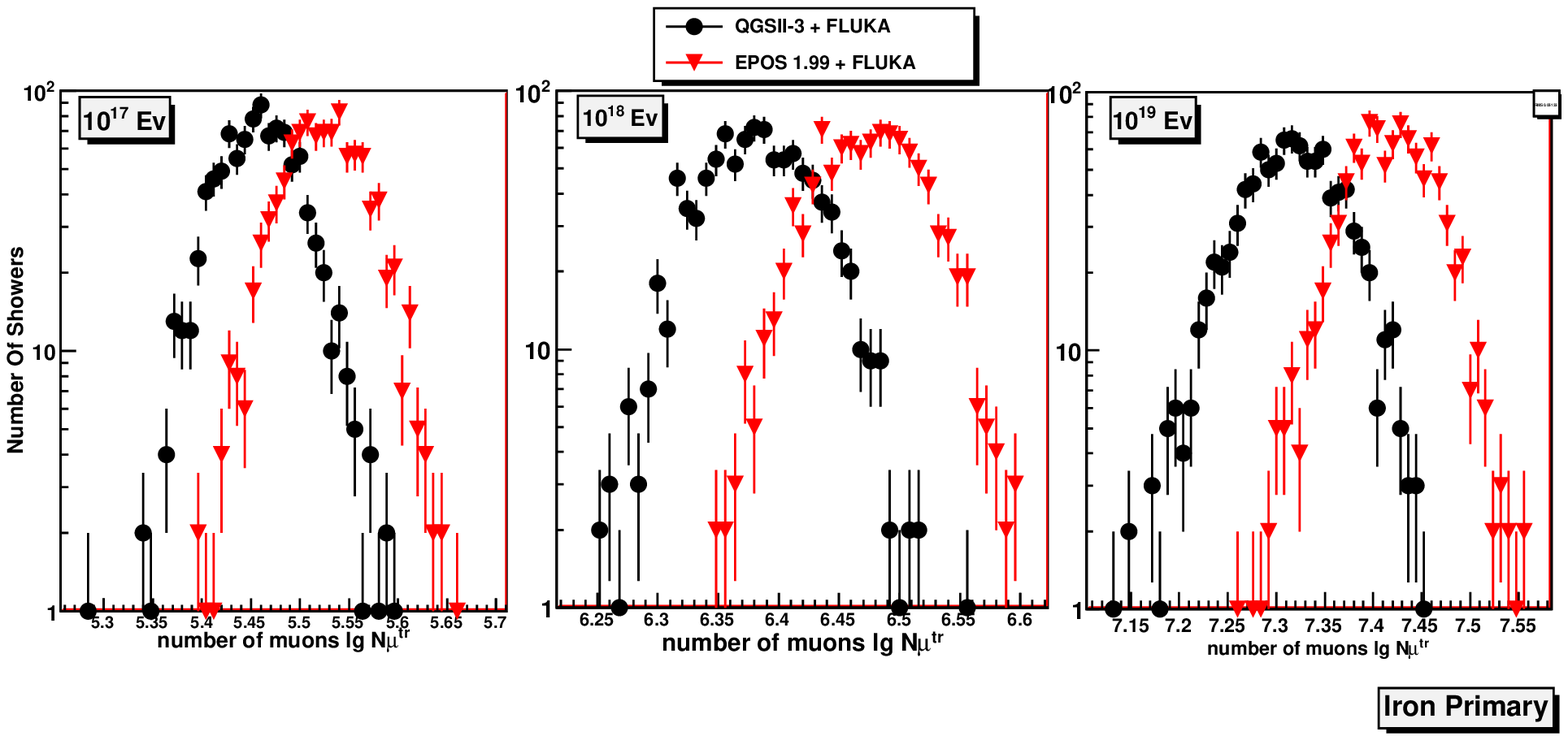}
\caption{Distribution of the number of truncated muons for iron primary}
 \label{Mu_iron}        
\end{figure} 
\end{center}

\begin{center}
\begin{table}
\caption{Average Truncated Muon Number and RMS value }
{\small
\resizebox {.5\textwidth }{!}{
\begin{tabular}{l| ll| ll }\hline
Energy/Model&\multicolumn{2}{c}{Proton Primary}&\multicolumn{2}{c}{Iron Primary}\\\cline{2-5}
$(eV)$&\multicolumn{1}{l}{$<lg(N_{\mu}^{tr})>$}&$RMS$&\multicolumn{1}{l}{$<lg(N_{\mu}^{tr})>$}&$RMS$\\\hline
$10^{17}-QGSII$&    5.33582&   0.09227&    5.45870&     0.04113\\
$10^{17}-EPOS$&     5.42664&   0.10664&    5.52427&    0.04210\\
$10^{18}-QGSII$&    6.26616&   0.09035&    6.38253&    0.04637\\
$10^{18}-EPOS$&     6.37158&   0.09563&    6.47433&    0.04434 \\
$10^{19}-QGSII$&    7.21209&   0.08441&    7.31510&     0.05116 \\ 
$10^{19}-EPOS$&     7.31632&   0.09665&    7.41770&     0.04526\\\hline
\end{tabular}
}}
\label{Mutable}
\end{table}
\end{center}

\section{Results}
The average of Depth of maximum \Xmax for the two models are plotted as function of energy 
in Figures \ref{X_max}. It is seen that heavier primary produces shower maximum at lower depth compared
with lighter primary as expected. The average total number of  electrons($N_e$), muons($N_\mu$), truncated muons($N_\mu^{tr}$),
hadrons ($E_h>100 GeV$), and sum of energies of all the hadrons($E_h>100 GeV$) registered at the ground level for the interaction
models EPOS 1.99 and QGSJET II-03 are plotted as a function of energy  in Figures \ref{Ne}, \ref{Nmu}, \ref{Ntrmu}, \ref{hadno},
and  \ref{sumhad_en} respectively.

It is seen that, both models yield a linear dependence of these components with energy in Log-Log scale.

\subsection{Dependence of \Xmax with primary Energy}
\label{sec:x_maxrmrs}
The value of \Xmax and also the slope of the curve for proton primary for EPOS 1.99 is slightly higher than that of  QGSJET II-03. 
Again, for iron primary also the slope of the curve has higher value for EPOS 1.99, while the values of \Xmax are nearly equal for 
both the models.   
   
The fitted values of the slope $ER_{10}$ and intercept $X_{init}$ are given in the Table \ref{xmaxtable}.

\begin{center}
\begin{figure}[h]
\includegraphics[width=0.5\textwidth]{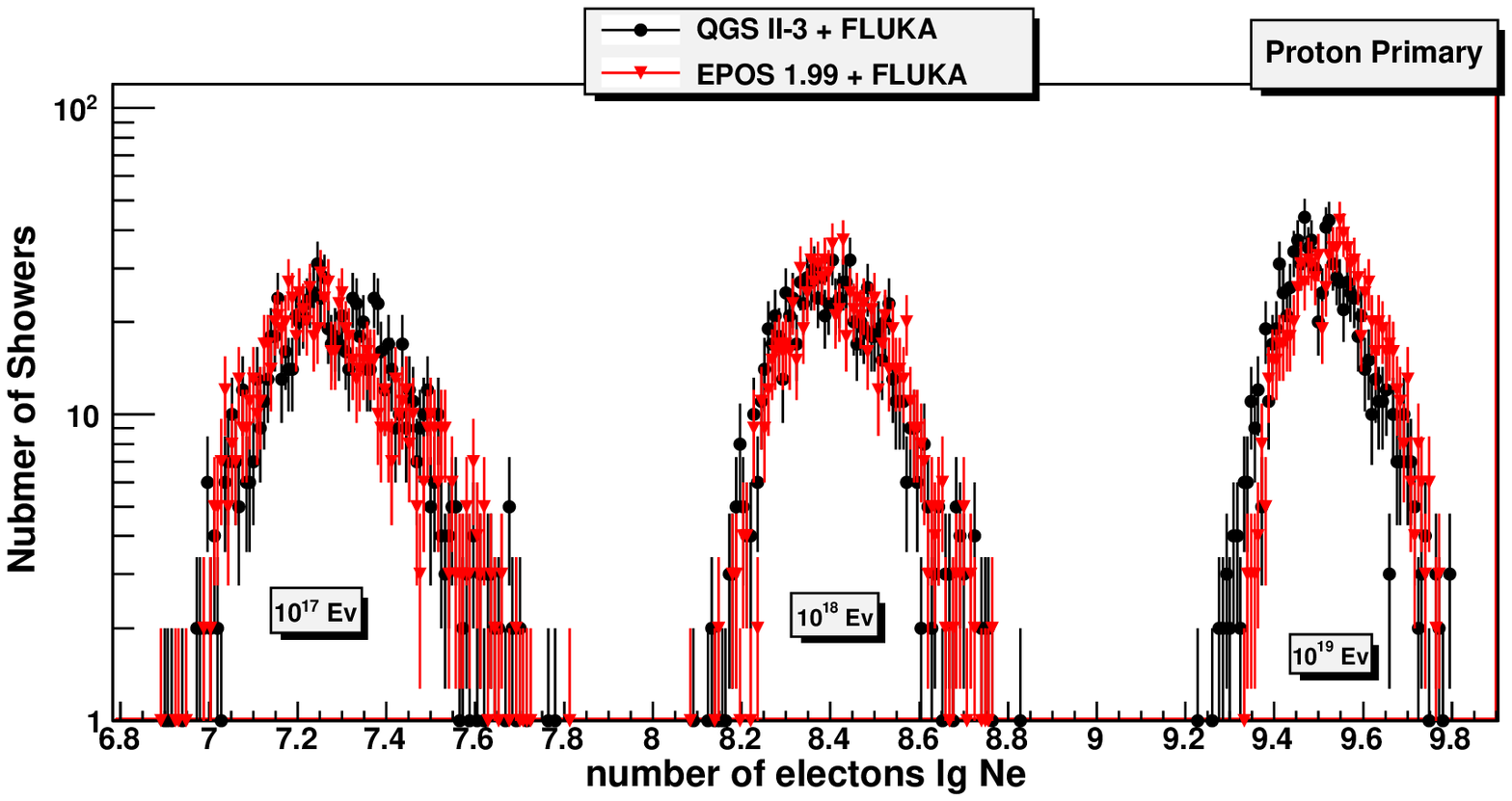}
\caption{Distribution of the number of electrons for proton primary  }
\label{El_Pr} 
\end{figure} 
\end{center}

\begin{center}
\begin{figure}[h]
\includegraphics[width=0.5\textwidth]{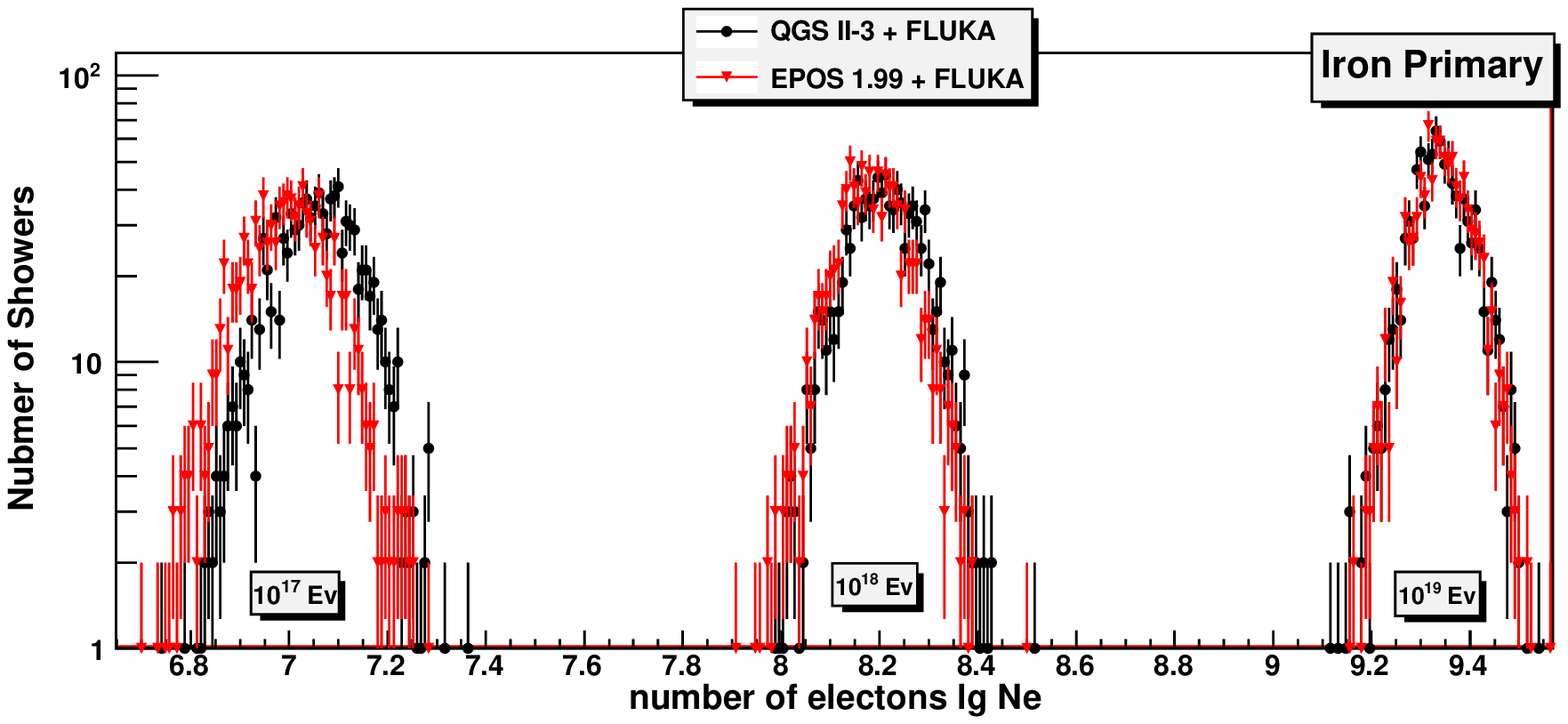}
\caption{Distribution of the number of electrons for iron primary  }
\label{El_Iron}
\end{figure} 
\end{center}

\begin{center}
\begin{table}
\caption{Average Electron Number and RMS value }
{\small
\resizebox {.5\textwidth }{!}{
\begin{tabular}{l|ll|ll}\hline
Energy/Model&\multicolumn{2}{c}{Proton Primary}&\multicolumn{2}{c}{Iron Primary}\\\cline{2-5}
$(eV)$&\multicolumn{1}{l}{$<lg(N_e)>$}&$RMS$&\multicolumn{1}{l}{$<lg(N_e)>$}&$RMS$\\\hline
$10^{17}-QGSII$&    7.28989&   0.14648&    7.05567&    0.08888\\
$10^{17}-EPOS$&     7.27924&   0.15111&    6.99535&    0.09171\\
$10^{18}-QGSII$&    8.40937&   0.11724&    8.21039&    0.07825\\
$10^{18}-EPOS$&     8.42408&   0.11274&    8.18449&    0.07497\\
$10^{19}-QGSII$&    9.50743&   0.09597&    9.34173&    0.06344\\ 
$10^{19}-EPOS$&     9.54035&   0.08953&    9.34330&    0.06105\\\hline
\end{tabular}
}}
\label{Eltable}
\end{table}
\end{center}

\subsection{Primary Energy Correlations}
\label{sec:size}
 From Figures \ref{Ne} through \ref{sumhad_en}, it is seen that, there is no significant differences in the electron shower
size and sum of hadron energy for the two models considered in the chosen primary energies for proton and iron primaries. 
However for muons (Fig. \ref{Nmu}, \ref{Ntrmu}) differences between the predictions of the two models are significant.
The number of muons is larger for EPOS1.99 model as compared to QGSJET II-03 for both the primaries. From the figure \ref{Nmu},
it is seen that there is very little overlap in the region bounded by p \& Fe primaries for two models considered.
However, for truncated muons (in the range 40\m-200\m) there is no overlap beyond $10^{18}$ {\it eV} (Figure \ref{Ntrmu}).       
For $ E_{h}>100GeV$ hadrons, total number for both proton \& iron primaries are found to be significantly more for 
EPOS than that for QGSJET II-3 (Figure \ref{hadno}).
However, considering the sum total of energies of all the hadrons ($ E_{h}>100GeV$), there is no significant difference
between the two models (Figure \ref{sumhad_en}).

\begin{center}
\begin{figure}[h]
\includegraphics[width=0.5\textwidth]{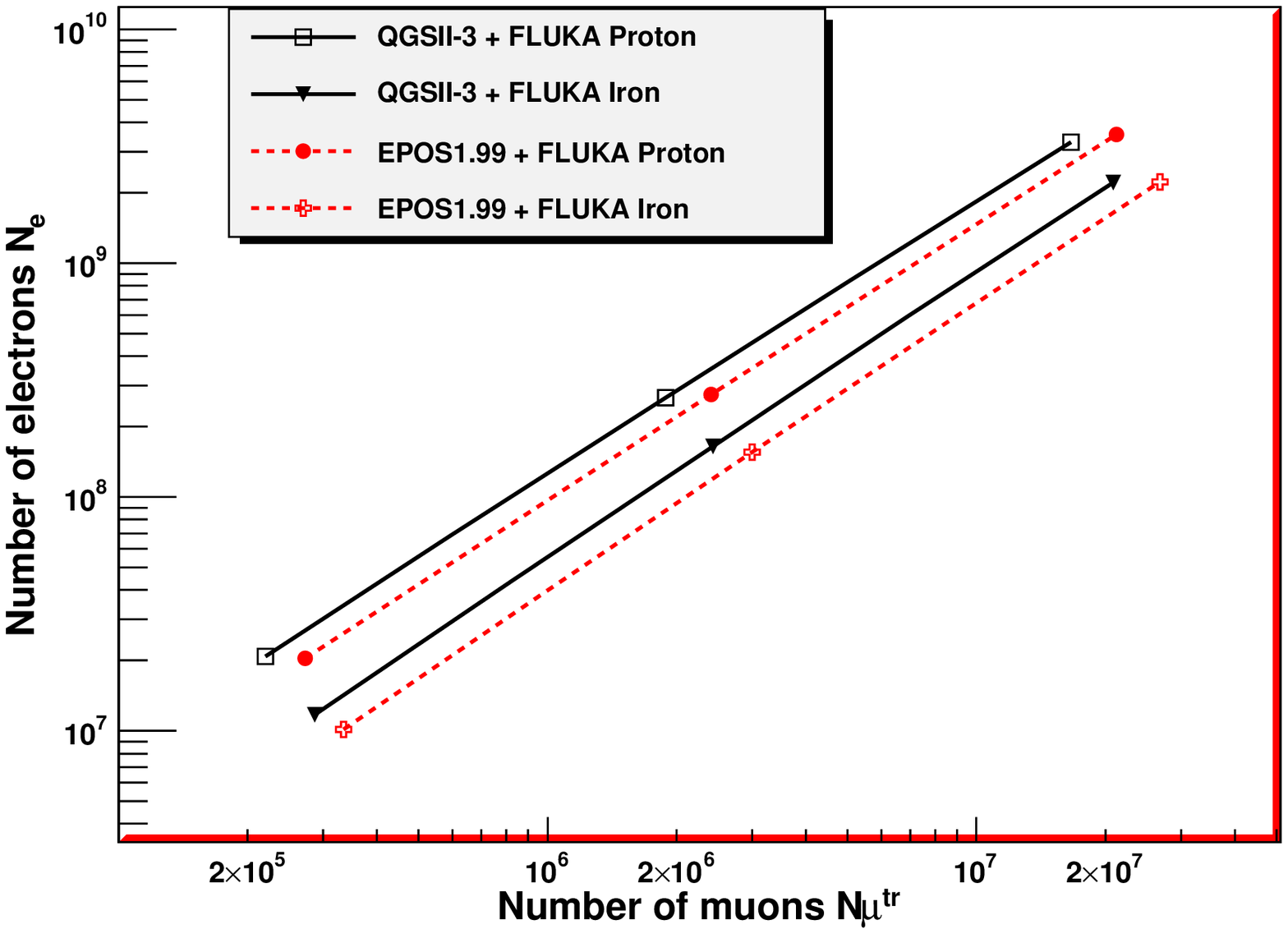}
\caption{Number of electrons Vs number of truncated muons  }
 \label{Tmu_el}         
\end{figure} 
\end{center}

\begin{center}
\begin{figure}[h]
\includegraphics[width=0.5\textwidth]{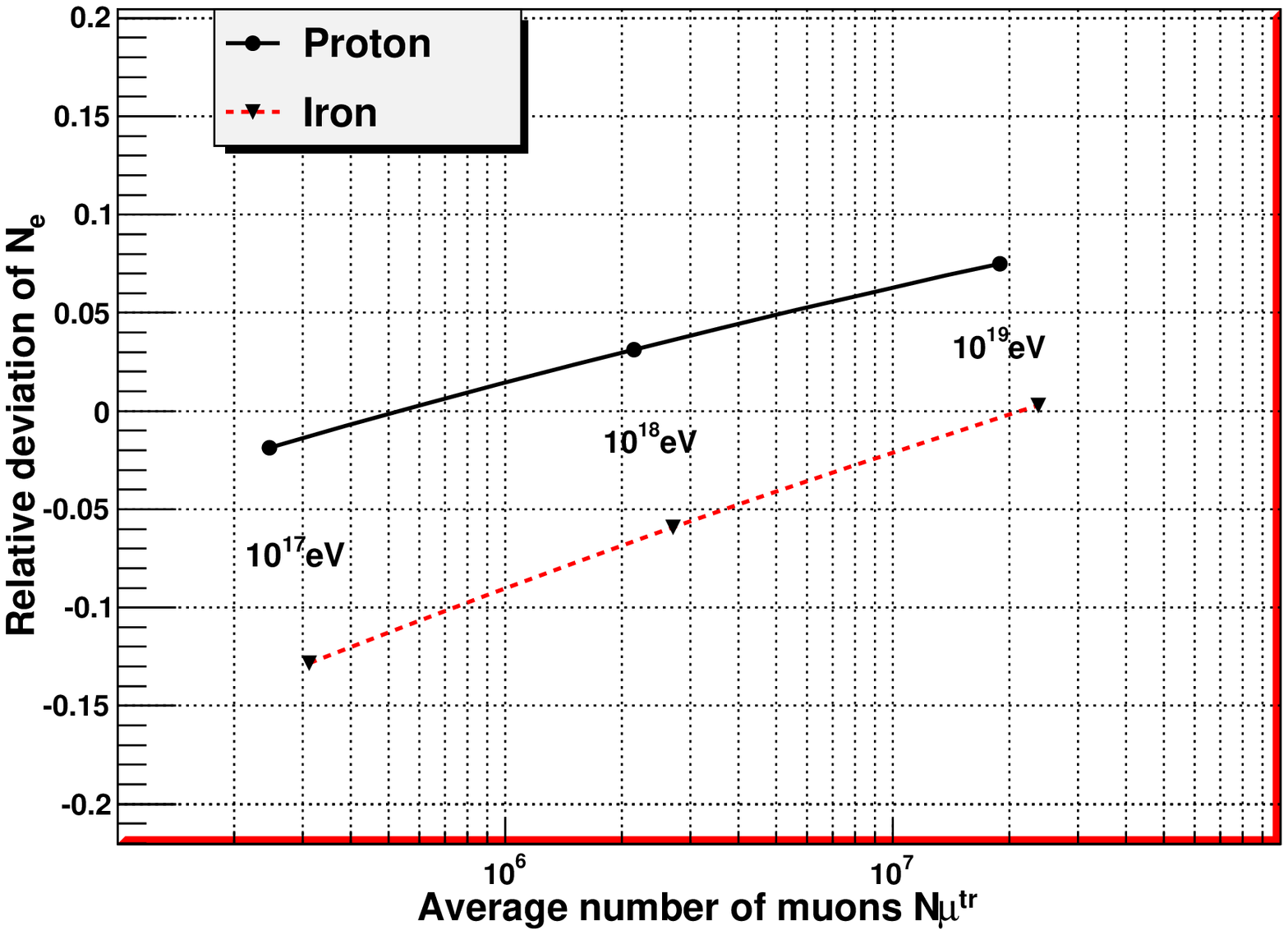}
\caption{Relative deviation of the number of electrons  Vs average number of truncated muons   }
 \label{Relel_mu}         
\end{figure} 
\end{center}

\subsection{Distribution of truncated muon numbers and electron numbers}
\label{sec:muel_dist}
   Distributions of truncated muon numbers $lg(N_{\mu}^{tr})$ for proton and iron  primaries are plotted in Figure
 \ref{Mu_Pr} and \ref{Mu_iron}. It is seen that EPOS1.99 yields higher value of $lg(N_{\mu}^{tr})$  than QGSJET II-03. 
The mean  \& s.d. of these distributions are tabulated in Table \ref{Mutable}, and significance test done (\ref{a3}). 

   Distributions of electron numbers $lg(N_e)$ for proton and iron  primaries are plotted in Figure
 \ref{El_Pr} and \ref{El_Iron}. It is seen that EPOS1.99 produces slightly less numbers of electrons 
for energy $10^{17}${\it eV} for iron primary, but at $10^{19}${\it eV}, EPOS1.99 produces slightly 
more electrons as tabulated in Table \ref{Eltable}, and significance test done(\ref{a3}).

\begin{center}
\begin{figure}[h]
\includegraphics[width=0.5\textwidth]{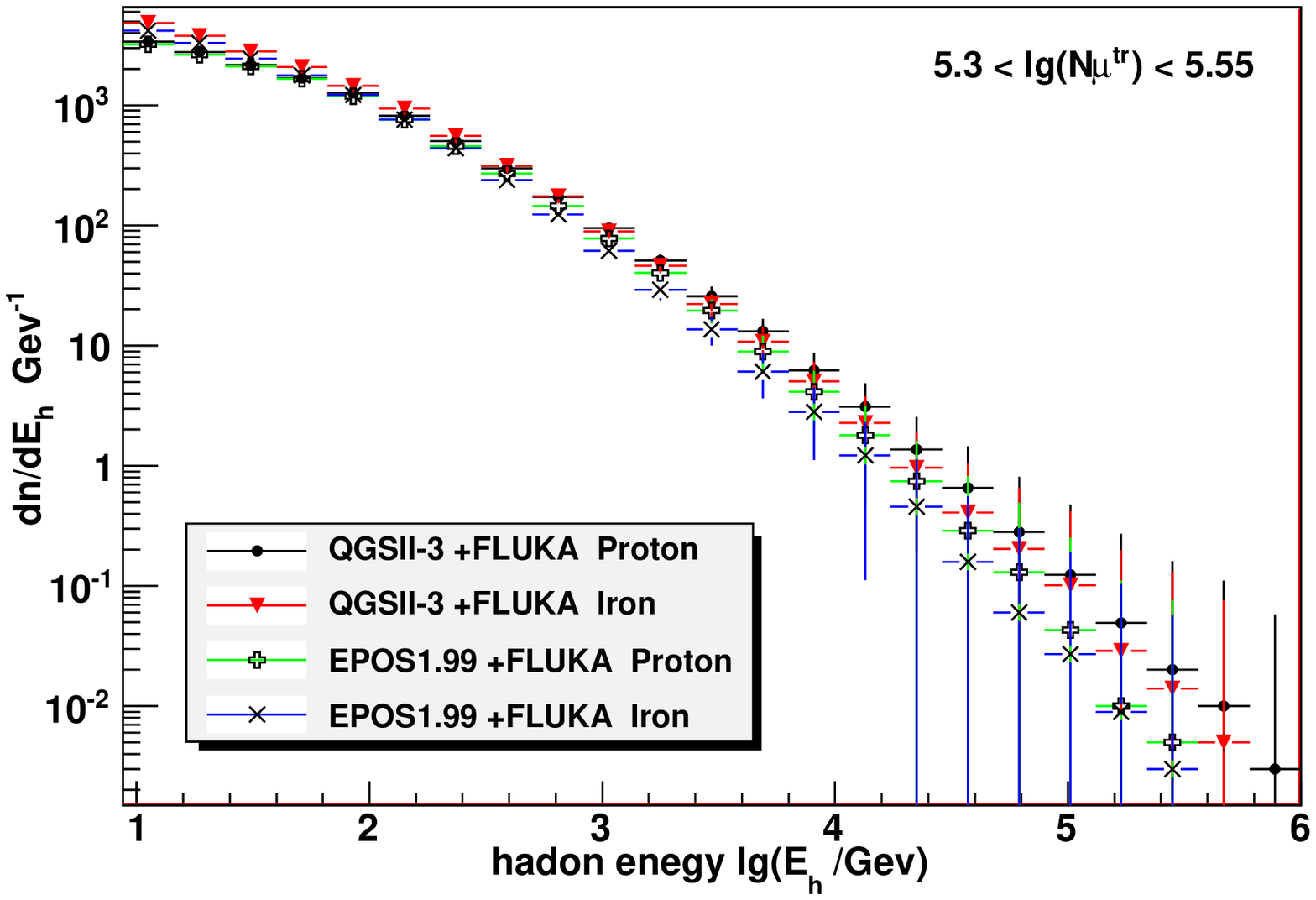}
\caption{hadron energy distribution $(10^{17} eV)$}
 \label{had_17}         
\end{figure} 
\end{center}

\begin{center}
\begin{figure}[h]
\includegraphics[width=0.5\textwidth]{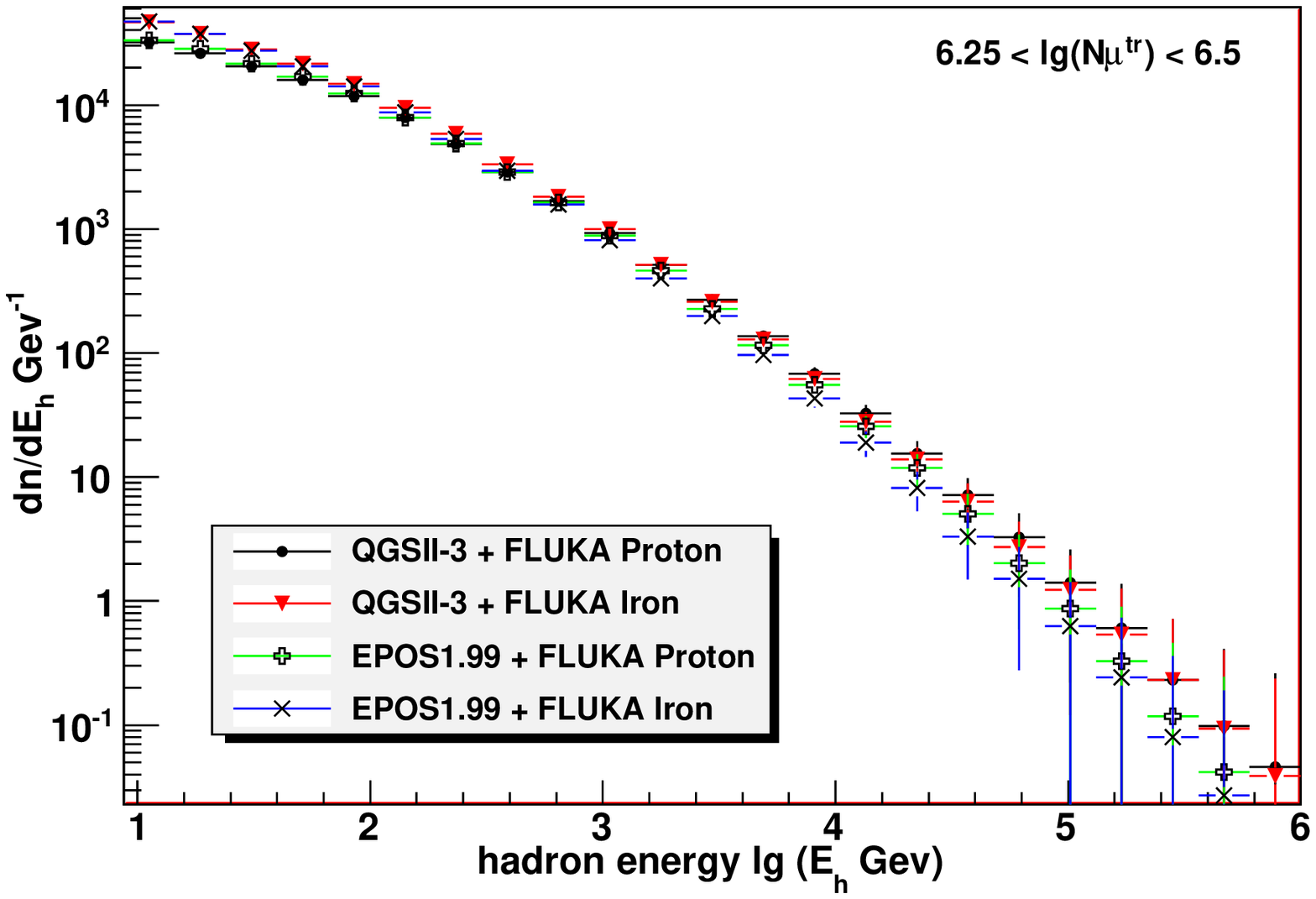}
\caption{hadron energy distribution $(10^{18} eV)$  }
 \label{had_18}         
\end{figure} 
\end{center}
  
   \subsection{Electron to muon correlation}
\label{sec:muel_corr}
     The average number of electrons as a function of the number of truncated muons are plotted in figure \ref{Tmu_el} 
for the two models. Slopes of all the curves are almost equal. To emphasize the differences between the model predictions,
relative deviation in the model prediction EPOS1.99 with respect to the QGSJET II-3 prediction, 
$(N_e^{EPOS} - N_e^{QGS})/N_e^{QGS}$  is plotted against the mean of $N_\mu^{QGS}$ and $N_\mu^{EPOS}$ in Figure \ref{Relel_mu}. 
It is seen that for proton primary, EPOS1.99 predicts slightly less (about 2\%) electrons for $10^{17}eV $ proton-induced 
showers, slightly more (about 3\%) electrons for $10^{18}eV $ and about 8\% more electrons for $10^{19}eV $ primary energy.
But for iron primary at  $10^{17}eV $,  $10^{18}eV $, and, $10^{19}eV $, EPOS 1.99 yields about 12\% lesser, about 6\% lesser
and nearly equal ( difference is less than 1\%)  numbers of electrons as compared with QGSJET II-3 predictions.

 \subsection{Variation in hadron component}
\label{sec:had_energy}
 
\begin{center}
\begin{figure}[h]
\includegraphics[width=0.5\textwidth]{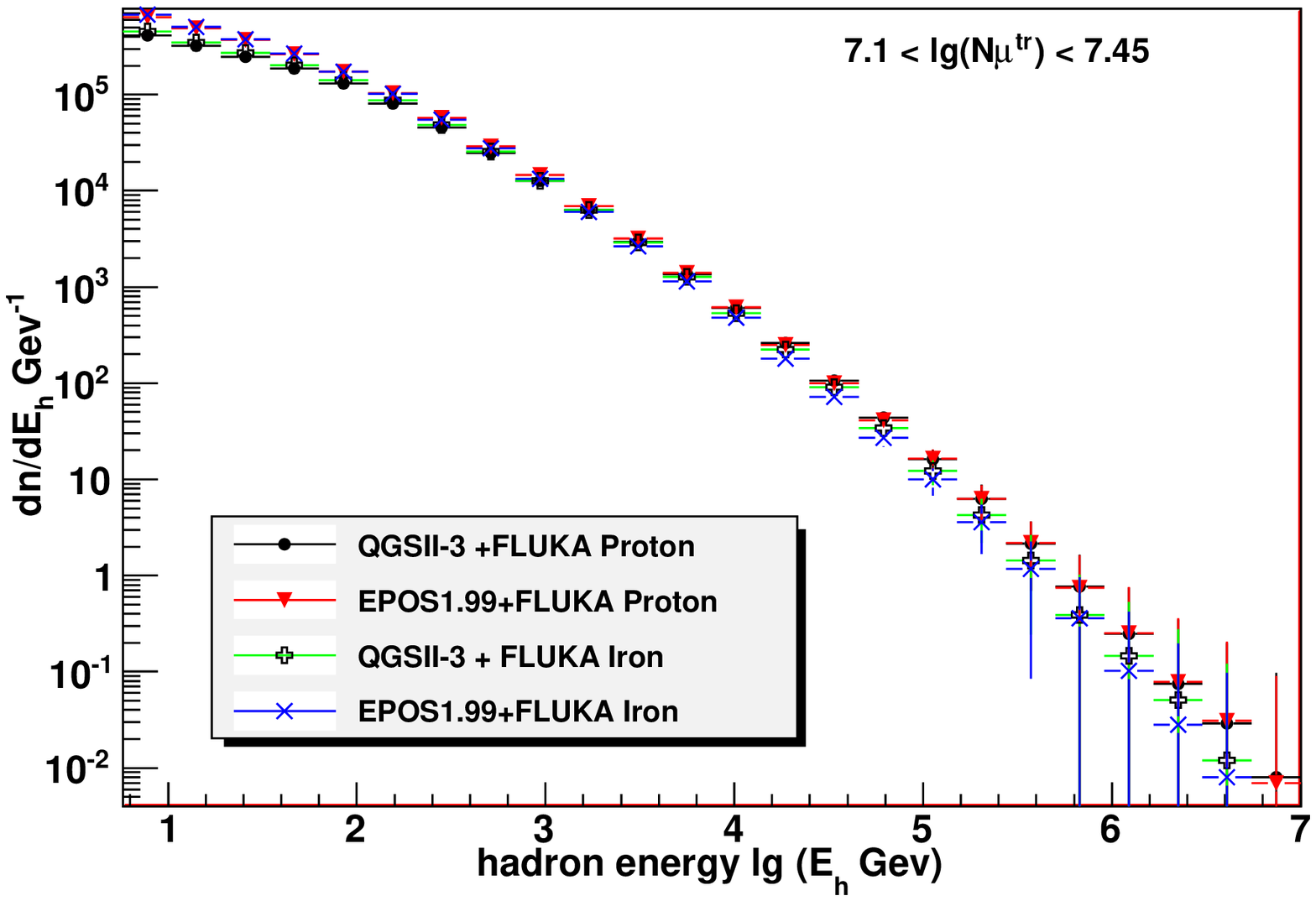}
\caption{hadron energy distribution $(10^{19} eV)$  }
 \label{had_19}         
\end{figure}
\end{center}

Total average number hadrons with energy $ E_{h}>100GeV$, for both primary masses \& at all the three energies show significant
difference between the two models (\ref{a4}). Energy spectrum of registered hadrons for $10^{17}eV $, $10^{18}eV $, and $10^{19}eV $ 
primary for both the models considered are displayed in the Figures \ref{had_17},\, \ref{had_18},\, and \ref{had_19} respectively. 
There are apparently no distinguishable differences. All the four data plots are overlapped to each-other with their error-bars.

\section{Summary}

Although \Xmax, shower size, hadron energy sum, hadron energy distribution shows no significant difference between the 
two HE models EPOS 1.99 (FLUKA) and QGSJET II-3 (FLUKA); muon number, hence electron to muon correlation shows incompatibility
between them. EPOS 1.99 (FLUKA) predicts more muons than QGSJET II-3 (FLUKA) which is known widely since its inception \cite{eposc}. 
This difference is explained as more (anti-)baryon production in EPOS leads to more muons in EAS \cite{eposc}. More (anti-)baryon 
generations in forward region result larger fraction of energy in the hadronic cascades and lesser in the electromagnetic 
cascades (more muon to electron ratio in the EAS) for EPOS prediction.
Also hadron($ E_{h}>100GeV$) shower size prediction by EPOS 1.99 (FLUKA) yields higher values irrespective of primaries 
considered as compared to that by QGSJET II-3 (FLUKA). Modifications may be needed for one, or, both of the models 
considered herein by comparing the  model predictions with experimental data.

\section{Acknowledgements}
The authors thankfully acknowledge UGC for computational infrastructure support under SAP  and C.C. Thakuria acknowledges 
UGC for fellowship under FDP.


\begin{appendix}
\section{Statistical Test}

\subsection{Z-test results for $X_{max}$ distribution (Table \ref{Zmaxtable})}
The null hypothesis is \\
$H_{0}$ = There is no difference between the two samples (QGSJET II-03 data and EPOS1.99 data).

\begin{center}
\begin{table}[h]
\caption{Z statistics for \Xmax }
{\small
\begin{tabular}{l|l|rl}\hline
Energy/Primary&\multicolumn{3}{c}{$X_{max} \, Dist^n$}\\\cline{2-4}
$(eV)$&\multicolumn{1}{c|}{$Z$}&$Inference$&\\\hline

$10^{17}-Proton$&    1.81&   Not &                Significant \\
                
$10^{17}-Iron$&      3.00&   Moderately &            Significant \\
                
$10^{18}-Proton$&    4.41&          &                Significant\\
                
$10^{18}-Iron$&     0.28&    Not &                 Significant\\
                
$10^{19}-Proton$&    8.65&   Highly &                Significant \\ 
              
$10^{19}-Iron$&      3.83&   Moderately &               Significant         \\\hline
\end{tabular}
  }
\label{Zmaxtable}
\end{table}
\end{center}

\subsection{Z-test results for $ER_{10}$ and $X_{init}$ (Table \ref{alphatable})}
The null hypothesis is \\
$H_{0}$ = There is no difference between the parameters for QGSJET II-03 and EPOS1.99 .

\begin{center}
\begin{table}[h]
\caption{Z statistics for $ER_{10}$ and $X_{init}$ }

\begin{tabular}{l|l|rl}\hline
Parameter &\multicolumn{3}{c}{$ER_{10}$ and $X_{init}$ }\\\cline{2-4}
$ \& Primary $&\multicolumn{1}{c|}{$Z$}& &$Inference$\\\hline

$ER_{10}-Proton$&    6.466&    &                Significant \\
                
$ER_{10}-Iron$&      4.981&     &                Significant \\
                
$X_{init}-Proton$&     5.933&     &                Significant\\
                
$X_{init}-Iron$&       4.834&     &                 Significant\\\hline
\end{tabular}
 
\label{alphatable}
\end{table}
\end{center}

It is seen that the parameters $ER_{10}$ and $X_{init}$ describing \Xmax (Equation 2), show
significant differences for EPOS and QGSJET II, irrespective of the primaries.

\subsection{Z-test results for $N_\mu^{tr}$ and $N_e$ distributions (Table \ref{Ztable})}
\label{a3}
The null hypothesis is \\
$H_{0}$ = There is no difference between the two samples (QGSJET II-03 data and EPOS1.99 data).
\begin{center}
\begin{table}[h]

\caption{Z statistics for $N_\mu^{tr}$  and $N_e$ dist$^N$s  }
{\small
\begin{tabular}{l|ll|ll}\hline
Energy/Primary&\multicolumn{2}{c}{$N_\mu^{tr} \, Dist^n$}&\multicolumn{2}{c}{$N_e \, Dist^n$}\\\cline{2-5}
$(eV)$&\multicolumn{1}{l}{$Z$}&$Inference$&\multicolumn{1}{l}{$Z$}&$Inference$\\\hline

$10^{17}-Proton$&    20.36&   Highly &                1.60&        Not \\
                &         &   Significant&                &       Significant \\
$10^{17}-Iron$&     35.22&    Highly &               14.93&       Highly \\
                &         &   Significant&                &       Significant \\
$10^{18}-Proton$&    25.34&   Highly &                2.86&       Moderately\\
                &         &   Significant&                &       Significant \\
$10^{18}-Iron$&     45.25&    Highly &                7.56&       Significant\\
                &         &   Significant&                &        \\
$10^{19}-Proton$&    25.68&   Highly &                7.93&       Significant \\ 
                &         &   Significant&                &         \\
$10^{19}-Iron$&     47.50&    Highly &                0.56&       Not         \\
                &         &   Significant&                &       Significant \\\hline
\end{tabular}
  }
\label{Ztable}
\end{table}
\end{center}

Average truncated muon number shows highly significant difference between the two models. For the average electron number, 
for lower energy($10^{17} eV$) high significance for iron primary, but no significant difference for proton primary is seen, whereas at 
higher energy($10^{19} eV$), no significance for iron, but significant differences for proton primary is observed ( Table \ref{Ztable}).

\subsection{Z-test results for total hadron number ($E_h>100GeV)$ (Table \ref{hadtable}) }
\label{a4}
The null hypothesis is 
$H_{0}$ = There is no difference between the two samples (QGSJET II-03 data and EPOS1.99 data).

\begin{center}
\begin{table}[h]
\caption{Z statistics for total hadron number ($E_h>100GeV)$ }
\begin{tabular}{l|l|rl}\hline
Energy/Primary&\multicolumn{3}{c}{$N_h \, Dist^n$}\\\cline{2-4}
$(eV)$&\multicolumn{1}{c|}{$Z$}&$Inference$&\\\hline

$10^{17}-Proton$&    4.07&          &                Significant \\
                
$10^{17}-Iron$&      21.09&   Highly &               Significant \\
                
$10^{18}-Proton$&    8.91&   Highly &                Significant\\
                
$10^{18}-Iron$&      6.52&    Highly &               Significant\\
                
$10^{19}-Proton$&    13.25&   Highly &               Significant \\ 
              
$10^{19}-Iron$&      15.59&   Highly &               Significant         \\\hline
\end{tabular}
 
\label{hadtable}
\end{table}
\end{center}
\end{appendix}

\end{document}